\documentstyle[epsf]{article}

\newcommand{\ct}{\cite}
\newcommand{\lb}{\label}

\newcommand{\bc}{\begin{center}}
\newcommand{\ec}{\end{center}}
\newcommand{\bd}{\begin{displaymath}}
\newcommand{\ed}{\end{displaymath}}
\newcommand{\be}{\begin{equation}}
\newcommand{\ee}{\end{equation}}
\newcommand{\ba}{\begin{array}}
\newcommand{\ea}{\end{array}}
\newcommand{\bt}{\begin{tabular}}
\newcommand{\et}{\end{tabular}}
\newcommand{\un}{\underline}
\newcommand{\ov}{\overline}
\newcommand{\bp}{\begin{picture}}
\newcommand{\ep}{\end{picture}}
\newcommand{\bfi}{\begin{figure}}
\newcommand{\efi}{\end{figure}}

\def\fun#1#2{\lower3.6pt\vbox{\baselineskip0pt\lineskip.9pt
\ialign{$\mathsurround=0pt#1\hfil##\hfil$\crcr#2\crcr\sim\crcr}}}

\begin{document}

%\vspace*{-1cm}
\title{
\vspace*{-1cm}
\hspace{3cm}
{\large Preprint CERN TH 99--275, September 6, 1999}\\
\hspace{5cm}
{\large Preprint NBI--HE--99--1}\\
\vspace{0.5cm}
\huge\bf
{ Phase Transition Couplings in the Higgsed Monopole Model }}
\author{\bf{ L.V.Laperashvili}\footnote{E-mails: laper@heron.itep.ru;
larisa@vxitep.itep.ru; laper@hetws3.nbi.dk}\\
\it Niels Bohr Institute,\\
\it DK-2100, Copenhagen, Denmark;\\
\it Institute of Theoretical and Experimental Physics,\\
\it B.Cheremushkinskaya 25, 117218 Moscow, Russia \\[0.2cm]
{\bf H.B.Nielsen}
\footnote{E-mails: hbech@nbivms.nbi.dk; Holger.Bech.Nielsen@cern.ch}\\
\it Niels Bohr Institute,\\
\it DK-2100, Copenhagen, Denmark;\\
\it CERN\\
\it CH 1211 Geneva 23\\
\it Switzerland}
\date{}
\maketitle
\vspace*{0.1cm}
{\large

\un{Corresponding author}\\
(temporary address from 01.08.1999 to 01.08.2000):\\
Prof.H.B.Nielsen,\\
Theory Division\\
CERN\\
CH 1211 Geneva 23\\
Switzerland\\
Telephone: +41 227678757\\
E-mail: Holger.Bech.Nielsen@cern.ch}

\thispagestyle{empty}

\newpage

\thispagestyle{empty}

\vspace{10cm}
\begin{abstract}
Using a one-loop approximation for the effective potential in the Higgs model
of electrodynamics for a charged scalar field, we argue for the existence of
a triple point for the renormalized (running) values of the selfinteraction
$\lambda$ and the "charge" $g$ given by $(\lambda_{run}, g^2) =
( -\frac{10}{9} \pi^2,\frac{4}{3}\sqrt{\frac{5}{3}}{\pi^2})
\approx(-11, 17)$.
Considering the beta-function as a typical quantity we estimate that
the one-loop approximation is valid with accuracy of deviations not
more than 30\% in the region of the parameters:
$0.2 \stackrel{<}{\sim }{\large \alpha, \tilde{\alpha}}
\stackrel{<}{\sim }1.35.$
The phase diagram given in the present paper
corresponds to the above-mentioned region of $\alpha,\;\tilde \alpha$.
Under the point of view that the Higgs particle is a monopole with a
magnetic charge $g$, the obtained electric fine structure constant
turns out to be $\alpha_{crit}\approx{0.18_5}$ by the Dirac relation. This
value is very close to the $\alpha_{crit}^{lat}\approx{0.20}$
which in a $U(1)$ lattice gauge theory corresponds to the phase
transition between the "Coulomb" and confinement phases.
Such a result is very encouraging for the idea of an approximate
"universality" (regularization independence) of gauge couplings at
the phase transition point. This idea was suggested by the authors
in their earlier papers.
\end{abstract}
\newpage
\pagenumbering{arabic}

\vspace{1cm}

\normalsize
\section{Introduction}

\vspace{0.51cm}

The Standard Model (SM) describes well all experimental results known today.
Most efforts to explain the SM are devoted to Grand Unification Theories
(GUTs). The supersymmetric extension of the SM consists of taking the SM
and adding the corresponding supersymmetric partners \ct{1}.
The precision of the LEP data allows us to extrapolate three running constants
$\alpha_{i} (\mu)$ of the SM (i=1,2,3 for U(1), SU(2), SU(3) groups)
to high energies with small errors and we are able to perform the
consistency checks of GUTs.

In the SM based on the group
\be
SMG=SU(3)_c\otimes SU(2)_L\otimes U(1)_Y  \lb{1}
\ee
the usual definitions of the coupling constants are used:
\be
\alpha_1=\frac{5}{3}\frac{\alpha}{\cos^2\theta_{\overline{MS}}},\quad
\alpha_2=\frac{\alpha}{\sin^2\theta_{\ov{MS}}},\quad
\alpha_3\equiv\alpha_S=\frac{g^2_S}{4\pi},
\lb{2}
\ee
where $\alpha $ and $\alpha_s$ are the electromagnetic and strong fine
structure constants, respectively. All of these couplings, as well as the
weak angle, are defined here in the Modified Minimal Subtraction
scheme $(\ov{MS})$ (see Reviews of Particle Physics).
Using experimentally given parameters and the renormalization
group equations (RGE), it is possible to extrapolate the experimental
values of three inverse running constants $\alpha_i^{-1}(\mu)$
to the Planck scale:
\be
\mu_{Pl} = 1.22\cdot {10}^{19}\; GeV.   \lb{3}
\ee
The comparison of the evolutions of the inverses of the running coupling
constants in the Minimal Standard Model (MSM) (with one Higgs doublet)
and in the Minimal Supersymmetric Standard Model (MSSM) (with two Higgs
doublets) shows the possibility of the existence of the grand unification
point at $\mu_{GUT}\sim 10^{16}$ GeV only in the case of the MSSM
(see Ref.\ct{2}).
But the absence of supersymmetric particle production at current
accelerators and additional constraints arising from limits on the
contributions of virtual supersymmetric particle exchange in a variety
of the SM processes indicate that at present there are no unambiguous
experimental results requiring the existence of supersymmetry.

Scenarios based on the Anti-Grand Unification Theory (AGUT) was developed
in Refs.\ct{3}-\ct{14} as a realistic alternative to SUSY GUTs
(see Ref.\ct{13}).
AGUT suggests the following assumption: supersymmetry does not exist
up to the Planck scale. There is no new physics (new particles,
superpartners) up to around an order of magnitude under this scale, and the
renormalization group extrapolation of experimentally determined couplings
to the Planck scale is contingent to not encountering new particles.

AGUT suggests that at the Planck scale $\mu_{Pl}$, considered as a fundamental
scale, there exists the more fundamental gauge group $G$, containing $N_{gen}$
copies of the Standard Model group $SMG$:
\be
G=SMG_1\otimes SMG_2\otimes \ldots\otimes
SMG_{N_{gen}}\equiv(SMG)^{N_{gen}},    \lb{4}
\ee
where the integer $N_{gen}$ designates the number of quark and lepton
generations.

SMG by definition is the following factor group:
\be
SMG = S(U(2)\times U(3)) = \frac{U(1)\times SU(2)\times SU(3)}
     {\{{(2\pi, -1^{2\times 2}, e^{i2\pi/3}1^{3\times 3})}^n|n\in
      Z\}}.  \lb{5}
\ee
If $N_{gen}=3$, then the fundamental gauge group G is:
\be
G=(SMG)^3=SMG_1\otimes SMG_2\otimes SMG_3 ,      \lb{6}
\ee
or the generalized G:
\be
   G = {(SMG)}^3\otimes U(1)_f     \lb{7}
\ee
which follows from the fitting of fermion masses (see Ref.\ct{11}).
The group $G = {(SMG)}^3\otimes U(1)_f$ is a maximal gauge transforming
(nontrivially) the 45 Weyl fermions of the SM (which it extends)
without unifying any of the irreducible representations of
the group of the latter. Anomalies are absent in this theory.

The AGUT approach was used in conjunction with
the Multiple Point Principle (MPP) proposed
several years ago by D.L.Bennett and H.B.Nielsen \ct{7}-\ct{9}.
Another name for the same principle is the "Maximally Degenerate Vacuum
Principle" (MDVP).
According to this principle, Nature seeks a special point -- the multiple
critical  point (MCP) -- where the group $G$ undergoes spontaneous
breakdown to the diagonal subgroup:
\be
G\to G_{diag.subgr.}=\left\{g, g, g\parallel g\in SMG\right\} \lb{8}
\ee
which is identified with the usual (low-energy) group SMG.

The idea of the MPP has its origin in the lattice investigations of
gauge theories. In particular, Monte Carlo simulations on the lattice
of U(1)--, SU(2)-- and  SU(3)-- gauge theories indicate the
existence of a triple point. Using theoretical corrections
to the Monte Carlo results on the lattice, it is possible to make slightly
more accurate predictions of AGUT for the SM fine-structure
constants. The MPP assumes that the SM gauge couplings do not unify and
predicts the following values of the fine structure constants
at the Planck scale in terms of the phase transition
(critical) couplings taken from the lattice gauge theories:
\be
 \alpha_i(\mu_{Pl}) = \frac{\alpha_i^{crit}}{N_{gen}}
                  = \frac{\alpha_i^{crit}}{3}
                            \lb{9}
\ee
for $i=2,3$ and
\be
 \alpha_1(\mu_{Pl}) = \frac{\alpha_1^{crit}}{{\frac 12}N_{gen}(N_{gen}+1)}
                    = \frac{\alpha_1^{crit}}{6}   \lb{10}
\ee
for U(1).

This means that at the Planck scale the fine structure constants
$\alpha_Y\equiv\frac{3}{5}\alpha_1$, $\alpha_2$ and $\alpha_3,$
as chosen by Nature, are just the ones corresponding to the Multiple
Critical Point (MCP) which is a point where all action parameter (coupling)
values meet in the phase diagram of the regularized  Yang-Mills $(SMG)^3$
-- gauge theory. Nature chooses coupling constant values
such that a number of vacuum states have the same energy density.
Then all (or just many) numbers of phases convene at the MCP and
different vacua are degenerate.

The extrapolation of the experimental values of the inverses
$\alpha^{-1}_{Y,2,3}(\mu)$ to the Planck scale $\mu_{Pl}$ by the
renormalization group formulae (under the assumption of a "desert"
in doing the extrapolation with one Higgs doublet) leads to the
following result:
\be
\alpha^{-1}_Y(\mu_{Pl})=55.5;\quad\alpha^{-1}_2(\mu_{Pl})=49.5;\quad\alpha
^{-1}_3(\mu_{Pl})=54.        \lb{11}
\ee
Using the AGUT prediction given by Eq.(\ref{10}) and the first value of
Eq.(\ref{11}) we have the following AGUT estimation for the U(1)
fine structure constant at the phase transition point:
\be
              \alpha_{crit}^{-1}\sim 9.    \lb{12}
\ee

Previously we have also speculated and made supportive calculations
\ct{12}-\ct{14}
that the dependence of the cut-off procedure is rather small:
phase transition coupling constants would not differ very much using
one or the other regularization. So we hope that there exists what
one might call "an approximate universality".

For the second-order phase transitions the exact universality would
be expected, but for the first-order phase transitions, which we really
hope for in the mentioned works, such an exact universality would
be quite unexpected. However, we still believe that an approximate one
really exists. Indeed, the main purpose of the present article is
to calculate the phase transition couplings and to confirm this
desired "approximate universality". The point is that the cut-off
considered in the previous works for MPP coupling calculations was
connected with the existence of artifact monopoles in the theory:
there are artifact monopoles in the lattice gauge theory and also
in the Wilson loop action model which we proposed \ct{12}.

The idea now is: instead of using the cut-off, we introduce physically
existing monopoles as fundamental fields into the theory. Then one may
not use the cut-off, but can rather think of any cut-off, although
essentially it should no longer matter. In other words, we consider a
theory with monopoles and look for a, or rather several, phase
transitions connected with the monopoles forming a condensate in
the vacuum. Writing such a monopole theory in the dual field formulation,
we deal with the usual Higgs model.

Below, using the Zwanziger formalism \ct{15}--\ct{17} for the dual
Abelian gauge theory describing the system with two (dual and non-dual)
gauge fields and both magnetic and electric charges (see Section 3),
we confirm in Section 6 a rather simple expression for the effective
potential in the one-loop approximation which was obtained for the Higgsed scalar
electrodynamics in Ref.\ct{18} (see also Ref.\ct{19}), and investigate
the phase structure of the Higgs model. But now the Higgs scalar field
is identified with the monopole field having magnetic charge $g$.
This means that the electric charge $e$ is connected to the formal charge $g$
of this Higgs field via the Dirac relation:
\be
      eg = 2\pi,                     \lb{13}
\ee
Then we can define the electric and magnetic fine structure constants:
\be
     \alpha = e^2/4\pi = \pi/g^2,\quad \quad
     \tilde \alpha = g^2/4\pi = \pi/e^2.      \lb{14}                                         \lb{alf}
\ee
In Section 5 we investigate the renormalization group equations for
$\alpha  $ and $\tilde {\alpha} $ in the case of
the existence of both charges and confirm the Dirac relation
for all renormalized effective coupling constants. Thus, for the
arbitrary scale we have the following relation:
\be
        \alpha \tilde{\alpha} = \frac 14    \lb{15}
\ee
which is used in Section 7 for the calculation of the critical
(phase transition) fine structure constants.

It seems that the region of parameters $\alpha$ and
$\tilde{\alpha}$ near the phase transition point
($\alpha_{crit} \approx{0.2}$ and ${\tilde{\alpha}}_{crit}\approx 1.25$)
obtained in the lattice investigations \ct{21}-\ct{23} of the U(1) gauge
theory allows us to consider the perturbation theory in both electric and
magnetic sectors with accuracy of deviations
$\stackrel{<}{\sim} 30\%$ (see Section 5).

In the present paper we aim to give the explanation of lattice
results and, what is more important, to show that the first-order phase
transition arises in the Higgsed monopole model already on the level of
the "improved" one-loop approximation (see Section 7), which
describes the phase transition found on the lattice with acceptable accuracy.
Thus, if the lattice phase transition coupling roughly coincides
with our Coleman--Weinberg model it cannot depend much on lattice details.
As it will be shown below, such a perturbation theory reproduces the lattice
result. Thus we get the suggested "approximate universality" even
for first-order phase transitions.

\section{Phase Transition Coupling in Lattice U(1) Gauge Theory (Compact QED)}

As was mentioned in Section 1, the idea of the MPP is based on
the lattice investigations of gauge theories.
In particular, Monte Carlo simulations of the U(1) gauge theory
described by the following lattice action:
\be
S = \sum_{\Box }[\beta\cos{\Theta (\Box )}+\gamma\cos{2\Theta (\Box )}]
                              \lb{16}
\ee
(here $\Theta (\Box )$ is the plaquette variable) indicate the existence of
the triple point \ct{21}-\ct{23} on the phase diagram shown in
Fig.1. From this triple point emanate three phase borders: the phase border
"1" separates the totally confining phase from the phase where only the
discrete subgroup $Z_{2}$ is confined; the phase border "2" separates the
latter phase from the totally Coulomb-like phase; and the phase border
"3" separates the totally confining and totally Coulomb-like phases.

In the speculative ideas which were proposed to provide a mechanism for the
MPP degeneracy of vacua in nature, one typically gets the prediction
that the phase transition is first order \ct{7}-\ct{9} - for instance,
if the world is in a state analogous to a microcanonical ensemble leading
to a mixture of phases.

In the Higgs monopole model considered in this paper you find a priori
formally first-order transitions, but it is possible to give some estimates
showing that this may not always be expected to be true at the end.
We shall leave this problem for the next paper.

In search for a lattice formulation of QED with a second order phase
transition, the simple Wilson action was generalized to Eq.(\ref{16})
by including a double charge term with coupling $\gamma $ \ct{20}
in the expectation that the phase transition would be driven
towards second-order at sufficiently small negative values of $\gamma $.

The recent lattice simulations of compact QED \ct{23}-\ct{31} 
have still not succeeded in agreement when they clarify the order of
the phase transition near $\beta = 1$. However, simulations on the hyper-torus,
up to $\gamma =$ - 0.4, revealed the reappearance of a double peak on large
enough lattices \ct{28}. In addition, for $\gamma $ between + 0.2 and - 0.4
the critical exponent $\nu $ has been found to decrease towards $\frac 14 $
(which corresponds to the first-order phase transition \ct{30})
with increasing lattice size for toroidal as well for spherical geometry.
In some cases stabilization of the latent heat has been observed.
Now we have rather strong indications that, at least in the region up to
$\gamma =$ - 0.4, the phase transition is of first order.

Fig.1 represents this situation showing, the tricritical point at some
negative value of the parameter $\gamma $. Of course, it remains
desirable to check this result on still larger lattices.

In Ref.\ct{21} the behaviour of the effective fine structure constant
$\alpha(\beta)$ (here $\beta=1/e_0^2$, and $e_0$ is the bare electric
charge) were investigated in the vicinity of the phase transition point
for the case $\gamma = 0$, and the following values of the fine structure
constants at the phase transition point was obtained:
\be
\alpha_{crit}^{lat}\approx{0.20},\quad
{\tilde \alpha}_{crit}^{lat}\approx{1.25},\quad\quad
\mbox{at}\quad\quad \beta_{crit}\approx{1.011}.
                              \lb{17}
\ee

Considering the Villain lattice action which corresponds to the extended
Wilson action (\ref{16}) with $\gamma \approx{-\; 0.22}$, the authors of
Ref.\ct{23} revised the values of the renormalized electrical coupling
$\alpha $ obtained in Ref.\ct{21} and presented :
\be
  \alpha^{lat} \approx {0.1836},\quad
{\tilde \alpha}^{lat} \approx {1.36}\quad \\
\mbox{ in the vicinity of the phase transition point}.
                         \lb{18}
\ee
The compact lattice QED is essentially related to the monopoles.
The phase transition "Coulomb - confinement" is known to be associated with
the condensation of magnetic monopoles \ct{23},\ct{29},\ct{32}-\ct{34}.
Monopole vacuum loops renormalize the fine structure constant $\alpha$
by an amount proportional to the susceptibility of the monopole gas \ct{34}.
The enhancement factor of this renormalization was estimated in Ref.\ct{12}:
\be
       K = \frac{\alpha_{crit}(\mbox{Monte Carlo})}{\alpha_{crit}
(\mbox{theory without monopoles})}\approx{1.66}.
                                                \lb{19}
\ee

The power law scaling behaviour of the monopole mass and condensate was
observed in Ref.\ct{23}. Using their results we have extracted the ratio
of the monopole mass to the monopole condensate $\rho = <0|\Phi|0>$ ($\Phi$
is a monopole field) in the confinement phase, a little bit away from the
phase transition point (but near to it):
\be
             \frac{m}{\rho}\approx{2.4}.        \lb{20}
\ee
By this "little bit away" we allude to the fact that $m/\rho $ varies rather
little except for $\beta$ being shorter than $\Delta \beta \approx{0.04}$ away 
from $\beta_{crit}$. That is, the ratio (\ref{20}) takes place
in the slowly varying region when approaching the phase transition.

In the lattice gauge theory monopoles are not physical objects:
they are lattice artifacts driven to infinite mass in the continuum limit.

Also in Ref.\ct{12}, instead of the lattice hypercubic regularization, we have
considered rather new regularization using non-local Wilson
loop action in the approximation of circular loops of radii $R\ge a$.
It was shown that the critical fine structure constant is rather
independent of the  regularization method. Its value is given by the
following expression:
\be
             \alpha_{crit}\approx{0.204},     \lb{21}
\ee
in correspondence with the Monte Carlo simulation result (\ref{17}),(\ref{18})
on the lattice. Such a phase transition coupling "universality" is needed
too much for the fine structure constant predictions claimed from the MPP.
But independently of the Multiple Point Model, the "approximate universality"
(if it takes place) is an important phenomenon for the phase transition
in any gauge field theory.

\section{The Zwanziger Formalism for the Abelian Gauge Theory with Electric
and Magnetic Charges}

In the lattice gauge theories and in the nonlocal Wilson loop model
mentioned in Section 2, monopoles are artifacts of the regularization.
Let us assume now a physically existing fundamental regulator.
The new idea is to consider monopoles as fundamental fields.
With the aim of confirming the universality of the critical
couplings in this case, we have investigated for a phase transition
the quantum field theory with electric and magnetic charges
considering monopoles as Higgs scalar particles.

A version of the local field theory of electrically and magnetically charged
particles is represented by Zwanziger formalism \ct{15},\ct{16}
(see also \ct{17}) which considers
two potentials $A_{\mu}(x)$ and $B_{\mu}(x)$ describing one physical
photon with two physical degrees of freedom.
Now and below we call this theory QEMD ("quantum
electromagnetodynamics").

In QEMD the total field system of the gauge, electrically ($\Psi$)
and magnetically ($\Phi$) charged fields is described by the partition
function which has the following form in Euclidean space:
\be
 Z = \int [DA][DB][D\Phi ][D\bar{\Phi}][D\Psi ][D\bar{\Psi}]e^{-S}
                                                         \lb{22}
\ee
where
\be
       S = S_{Zw}(A,B) + S_{gf} + S_e + S_m.
                                                       \lb{23}
\ee
The Zwanziger action $S_{Zw}(A,B)$ is given by:
$$
      S_{Zw}(A,B) = \int d^4x [\frac 12 {(n\cdot[\partial \wedge A])}^2 +
                  \frac 12 {(n\cdot[\partial \wedge B])}^2 +\\
$$
\be
     +\frac i2(n\cdot[\partial \wedge A])(n\cdot{[\partial \wedge B]}^*)
       - \frac i2(n\cdot[\partial \wedge B])(n\cdot{[\partial \wedge A]}^*)],
                                                         \lb{24}
\ee
where we have used the following designations:
$$
       {[A \wedge B]}_{\mu\nu} = A_{\mu}B_{\nu} - A_{\nu}B_{\mu},
\quad {(n\cdot[A \wedge B])}_{\mu} = n_{\nu}{(A \wedge B)}_{\nu\mu},\\
$$
\be
\quad {G}^*_{\mu\nu} = \frac 12\epsilon_{\mu\nu\lambda\rho}G_{\lambda\rho}.
                                           \lb{25}
\ee

The actions $S_e$ and $S_m$:
\be
           S_{e,m} = \int d^4x L_{e,m}(x)         \lb{26}
\ee
describe, respectively, the electrically and magnetically charged matter
fields, and $S_{gf}$ is the gauge-fixing action.

At the same time we present the generating functional with external sources
$J_{\mu}^{(A)}, J_{\mu}^{(B)}, \eta$ and $\omega$:
$$
        Z[J^{(A)}, J^{(B)}, \eta, \omega ] =
$$
\be
= \int [DA][DB][D\Phi ][D\bar{\Phi}][D\Psi ][D\bar{\Psi}] e^{ - S
       + (J^{(A)},A) + (J^{(B)},B) + (\bar{\eta},\Phi ) + (\bar{\Phi},\eta )
       + (\bar{\omega},\Psi ) + (\bar{\Psi},\omega )}
                                                          \lb{27}
\ee
where
\be
      (J,A) = \int d^4x J_{\mu}(x)A_{\mu}(x),\quad \mbox{and}\quad
      (\bar{\eta},\Phi ) = \int d^4x \bar{\eta}(x)\Phi (x),\quad \mbox{etc.}
                                                            \lb{28}
\ee
Let us consider now the Lagrangian $L_m$ describing the Higgs scalar monopole
field $\Phi(x)$ interacting with the dual gauge field $B_{\mu}(x)$:
\be
   L_m(x) = -\frac{1}{4}{(\partial \wedge B)}^2 + \frac 12{|{\tilde D}_{\mu}
\Phi |}^2 + U(\Phi),                                              \lb{29}
\ee
where
\be
       {\tilde D}_{\mu} = \partial_{\mu} - igB_{\mu}   \lb{30}
\ee
is a covariant derivative for the dual field;
\be
     U(\Phi) = \frac{1}{2}{\mu_m}^2 {|\Phi|}^2 + \frac{\lambda_m}{4}{|\Phi|}^4
                                   \lb{31}
\ee
is the Higgs potential for monopoles.

The complex scalar field:
\be
      \Phi = \phi + i\chi     \lb{32}
\ee
contains Higgs and Goldstone boson fields $\phi (x)$ and $\chi (x)$,
respectively.

Now we have a number of possibilities to describe electrically charged
fields (below we give the Lagrangian expressions in Minkowski space).
They can be:

a) fermions (electrons) described by the Dirac Lagrangian:
\be
      L_e = L_e^{(f)} = \bar{\Psi}{\gamma}_{\mu}(iD_{\mu} - \mu_e)\Psi,
                                               \lb{33}
\ee

b) Klein--Gordon (complex) scalars:
\be
       L_e = L_e^{(s)} = \frac 12 [{|D_{\mu}\Psi|}^2 - {\mu_e}^2{|\Psi|}^2],
                                                      \lb{34}
\ee

\hspace{-1mm} or c) Higgs scalars:
\be
      L_e = L_e^{(Hs)} = \frac 12 {|D_{\mu}\Psi|}^2 - U(|\Psi|)
                                                   \lb{35}
\ee
where
\be
       D_{\mu} = \partial_{\mu} - ieA_{\mu},              \lb{36}
\ee
and
\be
      U(|\Psi|) = \frac 12 \mu_e^2{|\Psi|}^2 + \frac{\lambda_e}4{|\Psi|}^4
                                                \lb{37}
\ee
is the Higgs potential for the electrically charged field.

Using the generating functional (\ref{27}) it is not difficult to
calculate the propagators of the fields considered in this model.

Three "bare" propagators of the gauge fields $A_{\mu}$ and $B_{\mu}$:
$$
  Q_{\mu\nu}^{0(A)} = < A_{\mu}A_{\nu} > = \frac{\delta^2Z[J^{(A)},J^{(B)},
           \eta ,\omega ]}{\delta J_{\mu}^{(A)}\delta J_{\nu}^{(A)}},\quad
$$
$$
  Q_{\mu\nu}^{0(B)} = < B_{\mu}B_{\nu} > = \frac{\delta^2Z[J^{(A)},J^{(B)}
           \eta ,\omega ]}{\delta J_{\mu}^{(B)}\delta J_{\nu}^{(B)}},
$$
\be
  Q_{\mu\nu}^{0(AB)} = < A_{\mu}B_{\nu} > = \frac{\delta^2Z[J^{(A)},J^{(B)},
           \eta ,\omega ]}{\delta J_{\mu}^{(A)}\delta J_{\nu}^{(B)}}
                                                             \lb{38}
\ee
were calculated by the authors of Ref.\ct{17} in the momentum space:
\be
      Q_{\mu\nu}^{0(A,B)}(q) = \frac 1{q^2}\bigl(\delta_{\mu\nu} +
  \frac{q^2 + {M_{A,B}}^2}{M^2_{A,B}}\frac{q_{\mu}q_{\nu}}{{(n\cdot q)}^2}
      - \frac 1{(n\cdot q)}(q_{\mu}n_{\nu} + q_{\nu}n_{\mu})\bigr),
                                                   \lb{39}
\ee
and
\be
     Q_{\mu\nu}^{0(AB)} = \frac {i}{q^2}\epsilon_{\mu\nu\rho\sigma}
              \frac{q_{\rho}n_{\sigma}}{(n\cdot q)}.
                                                  \lb{40}
\ee
The parameters $M_{(A,B)}$ are connected with the gauge. The gauge-fixed 
action chosen in Ref.\ct{17}:
\be
     S_{gf} = \int d^4x [\frac{M^2_A}2{(nA)}^2 + \frac{M^2_B}2{(nB)}^2]
                                                     \lb{41}
\ee
has no ghosts.

All Lagrangians (\ref{33})--(\ref{35}) have the interaction term
$j_{\mu}^eA_{\mu}$ where $j_{\mu}^e$ is
the electric current. The interactions in the Lagrangian
(\ref{29}) are given by $j_{\mu}^mB_{\mu}$ (here $j_{\mu}^m$ is the magnetic
current) as well as by the "seagull" term $g^2B_{\mu}B_{\mu}{\bar\Phi}\Phi$.
The equivalent seagull terms are present in the Lagrangians (\ref{34}) and
(\ref{35}). The interaction between the electric and magnetic charges is
carried out via the propagator $Q_{\mu\nu}^{(AB)}$.

\section{Dual Symmetry}

Duality is a symmetry appearing in free electromagnetism
as invariance of the free (static) Maxwell equations:
\be
 \bf{ \Delta}\cdot \vec{\bf B} = 0, \quad\quad
 \bf{ \Delta}\times \vec{\bf E} = 0,                 \lb{42}
\ee
\be
 \bf{ \Delta}\cdot \vec{\bf E} = 0, \quad\quad
 \bf{ \Delta}\times \vec{\bf B} = 0,                 \lb{43}
\ee
under the interchange of electric and magnetic fields:
\be
    \vec{\bf E} \to \vec{\bf B}, \quad\quad
        \vec{\bf B} \to  -\vec{\bf E}.                  \lb{44}
\ee
Letting
\be
      F = \partial\wedge A = - (\partial\wedge B)^{*},        \lb{45}
\ee
\be
      F^{*} = \partial\wedge B = (\partial\wedge A)^{*},        \lb{46}
\ee
it is easy to see that the following equations:
\be
    \partial_\lambda F_{\lambda\mu} = 0                   \lb{47}
\ee
with the Bianchi identity:
\be
    \partial_\lambda F_{\lambda\mu}^{*} = 0,                \lb{48}
\ee
equivalent to Eqs.(\ref{42}), (\ref{43}) are invariant under the Hodge
star operation on the field tensor:
\be
     F_{\mu\nu}^{*} = \frac 12 \epsilon_{\mu\nu\rho\sigma} F_{\rho\sigma}
                                                  \lb{49}
\ee
(here $F^{**} = - F$).

This Hodge star duality applied to the free Zwanziger Lagrangian (\ref{24})
leads to its invariance under the following duality transformations:
\be
   F \leftrightarrow F^{*},\quad\quad (\partial\wedge A)
\leftrightarrow (\partial \wedge B),
   \quad\quad (\partial\wedge A)^{*} \leftrightarrow
         - (\partial \wedge B)^{*}.
                                                \lb{50}
\ee
Introducing the interacting Maxwell equations:
\be
    \partial_\lambda F_{\lambda\mu} = j_{\mu}^e,              \lb{51}
\ee
\be
    \partial_\lambda F_{\lambda\mu}^{*} = j_{\mu}^m,            \lb{52}
\ee
with the local conservation laws for electric and magnetic charge:
\be
         \partial_{\mu}j_{\mu}^{e,m} = 0,            \lb{53}
\ee
we immediately see the invariance of these equations under the exchange
of the electric and magnetic fields (Hodge star duality) provided that
at the same time the electric and magnetic charges and currents (and
masses of the electrically and magnetically charged particles if they
are different) are also interchanged:
\be
     e \leftrightarrow g,
\quad\quad j_{\mu}^e \leftrightarrow j_{\mu}^m
                            \lb{54}
\ee
(and $\mu_e \leftrightarrow \mu_m$). We shall consider this symmetry as
the generalized duality.

The quantum field theory with electric $e_i$ and
magnetic $g_i$ charges is selfconsistent if both charges are quantized
according to the famous Dirac relation \ct{38}:
\be
            e_ig_j = 2\pi n_{ij}          \lb{55}
\ee
when $n_{ij}$ is an integer. Considering $n_{ij}=1$ in Eq.(\ref{55})
we obtain the Dirac quantization condition (\ref{13}) in terms of the
elementary electric and magnetic charges.

If the fundamental electric charge $e$ is so small that it corresponds to
the perturbative electric theory, then magnetic charges are large and
correspond to the strongly interacting magnetic theory, and vice versa.
But below we consider some small region of $e,g$ values (we hope that
it exists) which allows us to employ the perturbation theory in both the
electric and magnetic sectors.

When nontrivial dyons -- particles with both electric and magnetic
charges simultaneously -- are present, then the analogue of the
Dirac relation becomes a bit more complicated and it then reads:
\be
         e_ig_j - e_jg_i = 2\pi n_{ij}           \lb{56}
\ee
which is duality invariant (see for example the review \ct{39} and the
references there).

The relation (\ref{56}) has the name of the Dirac--Schwinger--Zwanziger
\ct{15},\ct{38},\ct{40} quantization condition. But the dyon theory
is not exploited in this paper.

Now we are ready to calculate the effective potential. But first we prefer
to consider the Dirac relation and the renormalization group equations
for the renormalized electric and magnetic fine structure constants.

\section{Renormalization Group Equations for the Electric and Magnetic
Fine Structure Constants. The Dirac Relation}

It is well known that in the absence of monopoles, the Gell-Mann--Low 
equation has the following form:
\be
     \frac{\mbox{d}(\log \alpha (p))}{\mbox{d}t} = \beta (\alpha (p))
                          \lb{57}
\ee
where
\be
          t = \log(p^2/M^2),   \lb{58}
\ee
$p$ is a 4-momentum and $\alpha = e^2/4\pi$.

The Gell-Mann--Low function $\beta(\alpha )$ depends on the Lagrangian
describing theory. As the was first shown in Refs.\ct{35},
at sufficiently small charge ($\alpha < 1$) the $\beta-$function
is given by series over $\alpha /4\pi$:
\be
\beta (\alpha ) = \beta_2 (\frac{\alpha}{4\pi}) +
                                  \beta_4 {(\frac {\alpha}{4\pi})}^2 + ...
                                            \lb{59}
\ee
The first two terms of this series were calculated in a QED long time ago
in Refs.\ct{35},\ct{36}.
The following result was obtained in the framework of the perturbation
theory (in the one- and two-loop approximations):

a)
\be
\beta_2 = \frac 43 \quad \mbox{and}\quad \beta_4 = 4 \quad
      -\quad \mbox{for fermion (electron) loops}
                                  \lb{60}
\ee
and

b)
\be
   \beta_2 = \frac 13 \quad \mbox{and}\quad \beta_4 =1 \quad
      -\quad \mbox{for scalar particle loops}.
                                          \lb{61}
\ee
This result means that for both cases a) and b)
the $\beta$-function can be represented by the following series
arising from Eq.(\ref{59}):
\be
  \beta(\alpha ) = \beta_2 (\frac{\alpha}{4\pi})( 1
                               + 3 \frac{\alpha}{4\pi} + ...)
                                                   \lb{62}
\ee
and we are able to exploit the one-loop approximation (given by
the first term of Eqs.(\ref{59}) and (\ref{62}))
up to $\alpha \sim 1$ (with accuracy $\approx 25\%$ for the
$\alpha \simeq 1$).

It is necessary to comment that three- and higher loop approximations
depend on the renormalization scheme. We do not discuss this problem
in the present paper.

The Dirac relation and the renormalization group equations (RGE)
for the electric and magnetic fine structure constants $\alpha $
and $\tilde \alpha $ were investigated in detail by the authors in the
recent paper \ct{37} where the same Zwanziger formalism was developed
for QEMD. The following result was obtained:
\be
  \frac{ d\log \alpha(p)}{dt} = -
  \frac{ d\log {\tilde \alpha}(p)}{dt} =
   \beta^{(e)}(\alpha ) - \beta^{(m)}(\tilde \alpha ). \lb{63}
\ee
It is easy to see that these RG-equations are in accordance with
the Dirac relation (\ref{15}) and with the generalized duality
considered in Section 4.

In Eq.(\ref{63}) the functions
$\beta^{(e)}(\alpha )$ and $\beta^{(m)}(\tilde \alpha )$
are described by the contributions of the electrically and magnetically
charged particle loops, respectively. Their analytical expressions
coincide with the usual well-known $\beta$--functions of QED given by
Eq.(\ref{59}), at least on the level of the two-loop approximation.

It is necessary to give some explanations how the result
(\ref{63}) was obtained.

J.Schwinger had shown \ct{40} that the Dirac relation (\ref{13})
is valid not only for the "bare" $e_0$ and $g_0$, but also for the
renormalized effective charges $e$ and $g$:
\be
              eg = e_0g_0 = 2\pi .
                                     \lb{64}
\ee

Eq.(\ref{64}) confirms the equality:
\be
 \frac{d\log \alpha(p)}{dt} = - \frac{d\log \tilde \alpha(p)}{dt}
                                           \lb{65}
\ee
and means that the Dirac relation is valid for all scales, e.g. for all $t$
in RGE (\ref{63}).

If the derivative ${d\log \alpha(p)}/dt$ in QEMD is also only a
function of the effective fine structure constants as in the
Gell-Mann--Low theory then we can write, in general, the following RGE:
\be
 \frac{d\log \alpha(p)}{dt} = \beta_1(\alpha ) +
                    \beta_2(\tilde \alpha)+ C ,     \lb{66}
\ee
\be
 \frac{d\log {\tilde \alpha}(p)}{dt} = {\tilde \beta}_1(\alpha ) +
                 {\tilde \beta}_2(\tilde \alpha)+ \tilde C .    \lb{67}
\ee
In Eqs.(\ref{66}),(\ref{67}) the terms containing the product
$\alpha \tilde \alpha $ are absent due to the Dirac relation (\ref{15}).
Applying to Eqs.(\ref{66}),(\ref{67})
the duality symmetry (the invariance under the interchange
$\alpha \leftrightarrow \tilde \alpha$) and using Eq.(\ref{65})
(which is the consequence of the Dirac relation)
it is not difficult to establish the following relations:
$$
   \beta_1(\alpha ) = - \beta_2(\alpha ) = - {\tilde \beta}_1(\alpha ) =
           {\tilde \beta}_2(\alpha ),\\
$$
\be
        C = \tilde C = 0.
                                \lb{68}
\ee
This result confirms the validity of RGE (\ref{63}) with
the function $\beta^{(m)}(\tilde \alpha )$
given by the same analytical expressions as the $\beta^{(e)}$--function
in $\alpha $.

From Eqs.(\ref{59}) and (\ref{62}) we see that it is possible
to consider the perturbation theory for $\beta^{(e)}(\alpha )$ and
$\beta^{(m)}(\tilde{\alpha})$ simultaneously if both $\alpha $ and
$\tilde{\alpha}$ are sufficiently small. Then the functions $\beta^{(e,m)}$
are given by the usual series similar to (\ref{59}) and calculated in
QED. If we are limited by the two-loop approximation, we have the following
equations (\ref{63}) for the cases a) and b):
\be
  \frac{ d\log \alpha(p)}{dt} = -
  \frac{ d\log {\tilde \alpha}(p)}{dt} =
   \beta_2 \frac{\alpha - \tilde \alpha }{4\pi}( 1 + 3\frac{\alpha
          + \tilde \alpha}{4\pi} + ....).
                      \lb{69}
\ee
It is not difficult to see that two first terms of this series
(one-loop and two-loop contributions) coincide with previous
results of the perturbative QED, but we have the difference on the level
of higher-order approximations when the monopole (electric particle)
loops begin to play a role in the electric (monopole) loops.

According to Eq.(\ref{69}) the one-loop approximation works with
an accuracy of deviations
$\stackrel{<}{\sim}30\%$ if both $\alpha $ and $\tilde {\alpha}$
obey the following requirement:
\be
  0.25 \stackrel{<}{\sim }{\large \alpha, \tilde{\alpha}}\stackrel{<}
                {\sim } 1.
                                         \lb{70}
\ee
For the compact (lattice) QED Eqs.(\ref{17}) and (\ref{18}) demonstrate that
$\alpha $ and $\tilde \alpha\;$, considering in the vicinity of the phase
transition point, almost coincide with the borders of the requirement
(\ref{70}) given by the perturbation theory for $\beta$--functions.
We can expect that these phase transition couplings may be described
by the one-loop approximation with accuracy not worse than $(30-50)\%$ although,
strictly speaking, we do not know the exact behaviour of the
asymptotic series (\ref{59}) or (\ref{69}).

\section{The Coleman--Weinberg Effective Potential for the Higgs Model
with Electric and Magnetic Charged Scalar Fields}

The effective potential in the Higgs model of electrodynamics for
a charged scalar field was calculated in the one-loop approximation for the
first time by the authors of Ref.\ct{18}. The general methods of the
calculation of the effective potential are given in Ref.\ct{19}.
Using these methods we have constructed the effective potential
(also in the one-loop approximation) for the theory with electric and
magnetic charges. Such a QEMD is described by the partition function
(\ref{22}) with the action $S$ containing the Zwanziger action (\ref{24}),
gauge fixing action (\ref{41}) and the actions (\ref{26}) for matter fields.
Monopoles are considered in this theory as Higgs scalar particles and
the corresponding Lagrangian is given by Eq.(\ref{29}). Electrically charged
fields can be described by the Lagrangians (\ref{33})-(\ref{35}).

Let us consider now the shifts:
\be
 \Phi (x) = \Phi_B + {\hat \Phi}(x),\quad\quad \Psi = \Psi_B + {\hat \Psi}(x)
                                         \lb{71}
\ee
with $\Phi_B$ and $\Psi_B$ as background fields and calculate the
following expression
for the partition function in the one-loop approximation:
$$
  Z = \int [DA][DB][D\hat{\Phi}][D\hat{\bar{\Phi}}][D\hat{\Psi}][D\hat
{\bar{\Psi}}]\times \\
$$
$$
\exp\{ - S(A,B,\Phi_B + \hat{\Phi},\Psi_B + \hat{\Psi})
- \int d^4x [\frac{\delta S(\Phi)}{\delta \Phi(x)}|_{\Phi=
\Phi_B}{\hat \Phi}(x) +
  \frac{\delta S(\Psi)}{\delta \Psi(x)}|_{\Psi=\Psi_B}{\hat \Psi}(x)]\}\\
$$
\be
=\exp\{ - F[\Phi_B, \Psi_B, g^2, e^2, \mu_m^2, \mu_e^2,
                                \lambda_m, \lambda_e]\}.
                          \lb{72}
\ee

Using the representation (\ref{32}) and writing a similar one for
the complex scalar field $\Psi $:
\be
             \Psi = \psi + i\zeta,           \lb{73}
\ee
we obtain the effective potential:
\be
  V_{eff} = F[\phi_B, \psi_B, g^2, e^2, \mu_m^2, \mu_e^2, \lambda_m,
                      \lambda_e]
                                        \lb{74}
\ee
given by the function $F$ of Eq.(\ref{72}) for the constant background
fields:
\be
  \Phi_B = \phi_B = \mbox{const},\quad\quad \Psi_B = \psi_B = \mbox{const}.
                                                  \lb{75}
\ee
Lagrangians considered in Section 3 indicate that the interaction
between the electric charges and monopoles appears in the vacuum diagrams
only on the level of the two-loop approximation and in higher
orders of the perturbative corrections to the classical potential.

Thus, in the one-loop approximation we have:
\be
           V_{eff} = V_{eff}^{(m)} + V_{eff}^{(e)}.      \lb{76}
\ee
The potential $V_{eff}^{(e)}$ following from the Lagrangian (\ref{35})
was calculated in the one-loop approximation by authors of Ref.\ct{18}.
The same expression takes place for $V_{eff}^{(m)}$. Using from now
the designations: $\mu = \mu_m,$ $\lambda = \lambda_m $ -- we can present
the following expression for $V_{eff}^{(m)}$ \ct{18},\ct{19}:
$$
 V_{eff}^{(m)} = \frac{\mu^2}{2} {\phi_B}^2 + \frac{\lambda}{4} {\phi_B}^4 +
$$
\be
\frac{1}{64\pi^2}[ 3g^4 {\phi_B}^4\log\frac{\phi_B^2}{M^2}
+ {(\mu^2 + 3\lambda {\phi_B}^2)}^2\log\frac{\mu^2 + 3\lambda
\phi_B^2}{M^2} + {(\mu^2 +\lambda \phi_B^2)}^2\log\frac{\mu^2
+ \lambda \phi_B^2}{M^2}]
                           \lb{77}
\ee
where $M$ is the cut-off scale.

The same expression (\ref{77}), but with $\mu =\mu_e$, $\lambda =\lambda_e $
and with $\psi_B$ instead of $\phi_B$ takes place for the $V_{eff}^{(e)}$.

The effective potential (\ref{74}) has several minima. Their position
depends on $e^2, g^2, \mu^2_{(e,m)}$ and $\lambda_{(e,m)}$.

It is easy to see that the first local minimum occurs at $\psi_B=0$ and
$\phi_B=0$ and corresponds to the so-called "symmetric phase" which is
the Coulomb-like phase in our description.

There exists only one vacuum $\psi_B=0$ for the Lagrangians (\ref{33}) and
(\ref{34}) if we describe the electric sector by the cases a) and b).
But in all cases our model is interested in the phase transition from the
Coulomb-like phase "$\psi_B = \phi_B = 0$" to the confinement phase
"$\psi_B = 0,\phi_B \neq 0$". Thus, in our investigation we have to use:
\be
      V_{eff}^{(e)} = 0 \quad \mbox{and}\quad
                                V_{eff} = V_{eff}^{(m)}.
                                   \lb{78}
\ee

Let us consider now the second local minimum at $\phi_B=\phi_0$.
We have the phase transition from the Coulomb-like phase to the
confinement phase if the second local minimum at $\phi_B=\phi_0$
is degenerate with the first local minimum at $\phi_0=0$
(see solid curve in Fig.2).

To use this one-loop approximation for the effective potential
calculation in the parameter combinations giving degenerate minima --
as we want -- really means that for the case when there is a compensation
between the classical (bare) and the one-loop terms, the latter are of
the same order as the first ones, and then the loop expansion is a priori
not reliable. But it could of course still be hoped that the accuracy of
one-loop corrections would be sufficiently good even in the case of the
cancellation. What we are looking for is not to know the sign of the
effective potential exactly in a region where we are close to the
shift of the sign, but rather to know when one effective potential goes
below zero as a function of the gauge coupling, say. The latter could
have a better chance of being to sufficient accuracy calculable.

\section{Calculation of the Critical Coupling in the Monopolic Model
of U(1) Gauge Theory}

The effective potential is given by the following expression equivalent to
Eq.(\ref{77}):
\be
V_{eff} = \frac{\mu^2_{run}}{2}\phi_B^2 + \frac{\lambda_{run}}{4}\phi_B^4
     + \frac{\mu^4}{64\pi^2}\log\frac{(\mu^2 + 3\lambda \phi_B^2)(\mu^2 +
                                                 \lambda \phi_B^2)}{M^4}
             \lb{79}
\ee
where $\lambda_{run}$ is the running self--interaction constant
given by the expression standing before $\phi_B^4$ in Eq.(\ref{77}):
\be
  \lambda_{run}(\phi_B^2)
   = \lambda + \frac{1}{16\pi^2} [ 3g^4\log \frac{\phi_B^2}{M^2}
   + 9{\lambda}^2\log\frac{\mu^2 + 3\lambda \phi_B^2}{M^2} +
     {\lambda}^2\log\frac{\mu^2 + \lambda\phi_B^2}{M^2}].   \lb{80}
\ee
The running squared mass of Higgsed monopoles also follows from
Eq.(\ref{77}):
\be
   \mu^2_{run}(\phi_B^2)
   = \mu^2 + \frac{\lambda\mu^2}{16\pi^2}[ 3\log\frac{\mu^2 +
   3\lambda \phi_B^2}{M^2} + \log\frac{\mu^2 + \lambda\phi_B^2}{M^2}].
                                  \lb{81}
\ee
The conditions for the degenerate vacua are given by the following
equations:
\be
            V_{eff}(\phi_0^2) = 0,     \lb{82}
\ee
\be
V'(\phi_0^2)=\frac{\partial V_{eff}}{\partial \phi_B^2}|_{\phi_B=\phi_0} = 0
                                    \lb{83}
\ee
with the inequality
\be
V''(\phi_0^2) = \frac{\partial^2 V_{eff}}{\partial \phi^2_B}|_{\phi_B=\phi_0}
              > 0.
                                \lb{84}
\ee
It is easy to obtain the solution of Eqs.(\ref{82}) and (\ref{83})
assuming that the last term in Eq.(\ref{79}) is small.
Neglecting the third term in Eq.(\ref{79}) we obtain:
\be
V_{eff}\approx \frac{\mu^2_{run}}{2}\phi_0^2 +
             \frac{\lambda_{run}}{4}\phi_0^4
                                              \lb{85}
\ee
and Eq.(\ref{82}) gives us the following relation:
\be
    \mu^2_{run}\approx{ - \frac{\lambda_{run}}{2}\phi_0^2}.
                                                            \lb{86}
\ee
Considering the derivative of $V_{eff}$ over $\phi_B^2$
we have:
\be
   V'(\phi_0^2) =
\frac 12\{\mu_{run}^2 + \lambda_{run}\phi_0^2 +
\frac{\partial \mu_{run}^2}{\partial \phi_B^2}|_{\phi_B=\phi_0}
\phi_0^2 + \frac{1}{2}\frac{\partial \lambda_{run}}
{\partial\phi_B^2}|_{\phi_B=\phi_0}\phi_0^4\}.
                                                          \lb{87}
\ee
In the following calculations we replace the "bare" constants
$\lambda $ and $\mu $ by $\lambda_{run}$ and $\mu_{run} $
assuming that only renormalized constants have a sense in the
field theory considered (it is natural to think that they will
appear in higher orders of perturbative corrections).
From now we have not only the one-loop approximation.

Using Eqs.(\ref{80}), (\ref{81}) and (\ref{86}) we obtain:
\be
V'(\phi_0^2) = \frac 14 \phi_0^2 (\frac 3{16\pi^2}g^4
+ \lambda_{run} + \frac 9{20} \frac {{\lambda_{run}}^2}{\pi^2}).
                                        \lb{88}
\ee
Now it is easy to find the solution of Eqs.(\ref{82})
and (\ref{83}):
\be
      g^4_{crit} = - \frac{4}{3}(4\pi^2 +
             \frac{9}{5}\lambda_{run})\lambda_{run}.
                                                  \lb{89}
\ee
The next step is the calculation of the second derivative of the effective
potential:
\be
 V''(\phi_0^2) = \frac {V'(\phi_0^2)}{\phi_0^2} + \frac 14 \phi_0^2
    \bigl(1 + \frac {18}{20\pi^2}\lambda_{run}\bigr)
    \frac{\partial \lambda_{run}}{\partial \phi_B^2}|_{\phi_B=\phi_0}.
                                                    \lb{90}
\ee
The requirement:
\be
 V''(\phi_0^2) = \frac{\partial^2
V_{eff}}{\partial\phi^2_B}|_{\phi_B=\phi_0} = 0    \lb{91}
\ee
gives us a triple point $A$ at the phase diagram shown in Fig.3.
Three phases -- Coulomb and two confining ones -- are present
in this diagram. The phase border "1" separates the phases $(confinement)_1$
and $(confinement)_2$. Here we have:
\vspace{1mm}
\bc
   $V_{eff}(\phi_0)=0\quad $, $V'(\phi_0)=0\quad $ and $\quad V''(\phi_0)<0$
\ec
-- for the $(confinement)_1$ phase.

\vspace{1mm}

And
\bc
 $V_{eff}(\phi_0^2)<0\quad $, $V'(\phi_0^2)=0\quad $ and
                       $\quad V''(\phi_0^2)>0$
\ec
-- for the $(confinement)_2$ phase.

\vspace{1mm}

The phase transition border "1" in Fig.3 corresponds to something similar to
the case presented in Fig.2 by the dashed curve, where we have two minima at
$\phi =\phi_1$ and $\phi =\phi_2$:
\be
       V_{eff}(\phi_1^2) = V_{eff}(\phi_2^2),    \lb{92}
\ee
\be
        {V'}_{eff}(\phi_1^2) = {V'}_{eff}(\phi_2^2) = 0. \lb{93}
\ee
The curve "1" in Fig.3 is calculated in the vicinity of the triple point A by
means of Eqs.(\ref{92}) and (\ref{93}) and is described by the
following expression:
\be
    g^4 = \frac{8}{3}(-\frac{7\pi^2}{9} +
                \frac{\lambda_{run}}{5})\lambda_{run}.
                                                       \lb{94}
\ee
The phase border "2" in Fig.3 separates the $(confinement)_2$ and
Coulomb-like phases. This border is given by the following equations:
\be
 V_{eff}(\phi_{1,2}^2) = 0;\quad\quad V'_{eff}(\phi_{1,2}^2) = 0
                                     \lb{95}
\ee
which coincide with Eqs.(\ref{82}) and (\ref{83}), but now in
the region of numerically smaller and still negative $\lambda_{run}$
we have more than two minima (see dashed curve in Fig.2). At first, we had
two minima at $\phi=0$ and $\phi=\phi_0$. Now we have three minima, but the
previous $\phi_0$--minimum has transformed analytically into a maximum.

The curve "3" in Fig.3 given by Eq.(\ref{89}) corresponds to the border
between the $(confinement)_1$ and Coulomb-like phases.

The solution of Eqs.(\ref{82}), (\ref{83}) and (\ref{91})
gives us the intersection of the curves (\ref{89}) and (\ref{94})
which determines the position of a triple point. This point A in Fig.3
is given by
\be
    \lambda_{run}^{(A)} = - \frac{10}{9}\pi^2     \lb{96}
\ee
and
\be
     {\bigl( g^{(A)}\bigr)}^2 =
   g^2_{crit}|_{\mbox{for}\;\lambda_{run}=\lambda^{(A)}_{ run}}
    = \frac{4}{3}\sqrt{\frac{5}{3}}\pi^2\approx{17}.
                                \lb{97}
\ee
The last result follows from Eq.(\ref{89}) and corresponds to the
following triple point value of the magnetic fine structure constant:
\be
          \tilde \alpha^{(A)}\approx 1.35.          \lb{98}
\ee
Then the Dirac relation (\ref{15}) allows us to calculate the value of the
triple point electric fine structure constant:
\be
      \alpha^{(A)} = \frac{\pi}{{(g^{(A)})}^2}
                = \frac3{4\pi}\sqrt{\frac 35}\approx{0.18_5},
                                          \lb{99}
\ee
in agreement with the Monte Carlo lattice results (\ref{17}),(\ref{18})
and with the Wilson loop action model given by Eq.(\ref{21}).
Here we are successful in the confirmation of the
critical coupling approximate universality.

Notice that from Eq.(\ref{86}) we have $|\mu_{run}|>|\phi_0|$.
The estimation of their ratio at the triple point A gives:
\be
    \frac{|\mu_{run}|}{|\phi_0|}\{\mbox at\;the\;triple\;point\;A\}
          \approx \sqrt{\frac{\lambda_{run}^{(A)}}{2}}
          \approx 2.35.                \lb{100}
\ee
This value coincides with the lattice result (\ref{20}) with high accuracy.
Taking into account that our Higgsed monopole model gives the first-order
phase transition, it is a small wonder that we got the result corresponding
to the lattice confinement phase. At least, it is necessary to think about 
such coincidences.

Eq.(\ref{100}) shows that the mass of the monopole
is large compared to the $\phi_0$ scale at
which our calculation leading to the result (\ref{99}) is presumably correct.
So there are no RG running due to monopoles between
the $\phi_0$--scale and the infra-red limit. In consequence, the infra-red
limit coupling should not be RG corrected by monopole contributions and
will just be our value (\ref{99}).

Let us consider now if the assumption that the third term in
Eq.(\ref{79}) is negligibly small is selfconsistent with our
calculations.

We have the following expression equivalent to Eq.(\ref{79}):
\be
V_{eff} = \frac{\lambda_{run}}{4}\phi_B^4\{1 + \frac{2\mu^2_{run}}
{\lambda_{run}\phi_B^2} +
\frac{\mu^4}{16\pi^2\lambda_{run}\phi_B^4}\log\frac
{(\mu^2 + 3\lambda\phi_B^2)(\mu^2 + \lambda\phi_B^2)}{M^4}\}.
                                    \lb{101}
\ee
The largest value of $\mu^2$ is $\mu^2\approx{M^2}$,
which allows us to estimate
the value of the third term in the brackets of Eq.(\ref{101})
in the vicinity of the second minimum when $\phi_B=\phi_0$.
Using the designation:
\be
 V_{eff}^{(3)}(\phi^2)=\frac{\mu^4}{64\pi^2}\log{\frac{(\mu^2+3\lambda\phi^2)
(\mu^2+\lambda\phi^2)}{M^4}}
                                    \lb{102}
\ee
we can consider the ratio
\be
    R = V_{eff}^{(3)}(\phi_0^2)/({\frac{\lambda_{run}}{4}\phi_0^4})\approx
{\frac{{\mu_{run}}^4}{16\pi^2\lambda_{run}\phi_0^4}\log\frac
{(\mu^2+3\lambda\phi_0^2)(\mu^2+\lambda\phi_0^2)}{M^4}}.
                                                          \lb{103}
\ee
Then Eq.(\ref{86}) gives us:
\be
     R = \frac{\lambda_{run}}{64\pi^2}\log \frac
{(\mu^2+3\lambda\phi_0^2)(\mu^2+\lambda\phi_0^2)}{M^4}
              \stackrel{<}{\sim}\frac{\lambda_{run}}{64\pi^2}\log 5 .
                                              \lb{104}
\ee
At the triple point A the value of $R$ is determined by
Eq.(\ref{96}):
\be
     R^{(A)}\stackrel{<}{\sim}{0.018\log{5}}\approx{0.03}.
                                                      \lb{105}
\ee
Really, we have $R<<1$. This result confirms the negligibility
of the third term in Eq.(\ref{79}) for $V_{eff}$ near the
second minimum shown in Fig.2.

The values (\ref{99}) and (\ref{98}) obtained in our Higgsed
monopole model for the triple point electric and magnetic fine
structure constants correspond to the one-loop approximation for
the effective potential and can be improved by the consideration of
the higher-loop corrections. However, this result is close to
the borders of the perturbation theory requirement (\ref{70}).
The phase diagram shown in Fig.3 resembles the region (\ref{70})
having slight deviations.
If one adds electrically charged particles then the corrections from
their contributions are to be taken into account. But on the level of
the one-loop approximation we have only monopoles.
We think that in all cases our results are guaranteed with accuracy
less than $50\%$.
However, it seems that the idea of the
approximate universality of the critical coupling constants is
confirmed.

\section{Conclusions}

We have used the Coleman--Weinberg effective potential for the Higgs model
with the Higgs field $\phi$ conceived as a monopole scalar field
to enumerate a phase diagram suggesting that in addition to the phase with
$<\phi>=0$ (i.e. the Coulomb phase)
we have two different phases with $<\phi>\neq 0$
meaning, two different confinement phases $(confinement)_1$ and
$(confinement)_2$. These three phases meet
at a triple point and we calculated what we called the effective or
running $\lambda_{run}$ and $g^2$ couplings at this triple point A:
\be
  \bigl(\lambda^{(A)}_{run},{(g^{(A)})}^2\bigr) = ( -\frac{10}{9}\pi^2,
\frac43\sqrt{\frac53}\pi^2)\approx{(-11,17)}.  \lb{106}
\ee
By the Dirac relation this calculated ${(g^{(A)})}^2$ corresponds to
\be
     \alpha^{(A)} = \frac{\pi}{{(g^{(A)})}^2} = \frac{3}{4\pi}\sqrt{\frac 35}
      \approx{0.18_5}.                         \lb{107}
\ee
and
\be
     {\tilde \alpha}^{(A)}\approx 1.35.          \lb{108}
\ee

It is noticed that these triple point fine structure constant values
coincide rather well with the values of the fine structure constant
at the phase transition point for a U(1) lattice gauge theory
(see Eqs.(\ref{17}) and (\ref{18})).
But for the values (\ref{107}) and (\ref{108}) giving the perturbative region
of parameters:
\be
0.2 \stackrel{<}{\sim }{\large \alpha, \tilde{\alpha}}
\stackrel{<}{\sim }1.35    \lb{109}
\ee
we cannot guarantee the accuracy of deviations better than $30\%$,
as follows from the estimation of the two-loop contributions
(Section 5).

Hereby we see a strong argument for our previously hoped-for principle
of "approximate universality" for the first-order phase transitions:
the fine structure constant (in the continuum) is at the/a multiple
point approximately the same one, independent of various parameters
of the (lattice e.g.) regularization.

This is indeed first suggested by the agreement of the above-obtained
value $\alpha^{(A)}=0.18_5$ with the phase-border value in the various
different regularizations. Secondly we could also argue:

(All) various different regularizations for U(1) electrodynamics which
usually should have artifact monopoles could, in the philosophy of going
back and forth  between continuum and, say, lattice regularization, be
described by the Higgs model with $\phi$ interpreted as a monopole
scalar field.

Since we showed that we could calculate the triple point in this
continuum theory all the various regularizations in which artifact
monopoles are presumed connected with the phase transitions must have
approximately the same continuum parameters at the triple point.

All different versions of U(1) lattice gauge theories have normally
artifact monopoles. If they are approximated by a continuum field
model it should be the Higgs model interpreted as in the present article
and our triple point $\alpha^{(A)}\approx 0.18_5$ would be the coupling
at the triple point of $\un{whatever}$ U(1) lattice gauge theory. This is our
previously suggested "approximate universality" which is quite necessary
for the AGUT and MPP predictions. To the point, the result (\ref{107})
obtained in our Higgsed monopole model gives:
\be
          \alpha_{crit}^{-1}\approx 5.4  \lb{110}
\ee
which is comparable (in the framework of our accuracy) with AGUT-MPP
prediction (\ref{12}).
The details of this problem are discussed in Refs.\ct{7}-\ct{9}.

We have a hope that the two-loop approximation corrections to the
Coleman--Weinberg effective potential will lead to much better accuracy
in the calculation of the phase transition couplings, but this is an
aim of our next papers.
\vspace*{0.5cm}

ACKNOWLEDGMENTS: We would like to express a special thanks to
D.L.Bennett for useful discussions, P.A.Kovalenko, D.A.Ryzhikh and Yasutaka
Takanishi for help.
We are also very thankful to
Colin Froggatt and Ivan Shushpanov for stimulating interactions.
Financial support from grants INTAS-93-3316-ext and INTAS-RFBR-96-0567
is gratefully acknowledged.

\newpage
\begin{figure}
\centerline{\epsfxsize=\textwidth \epsfbox{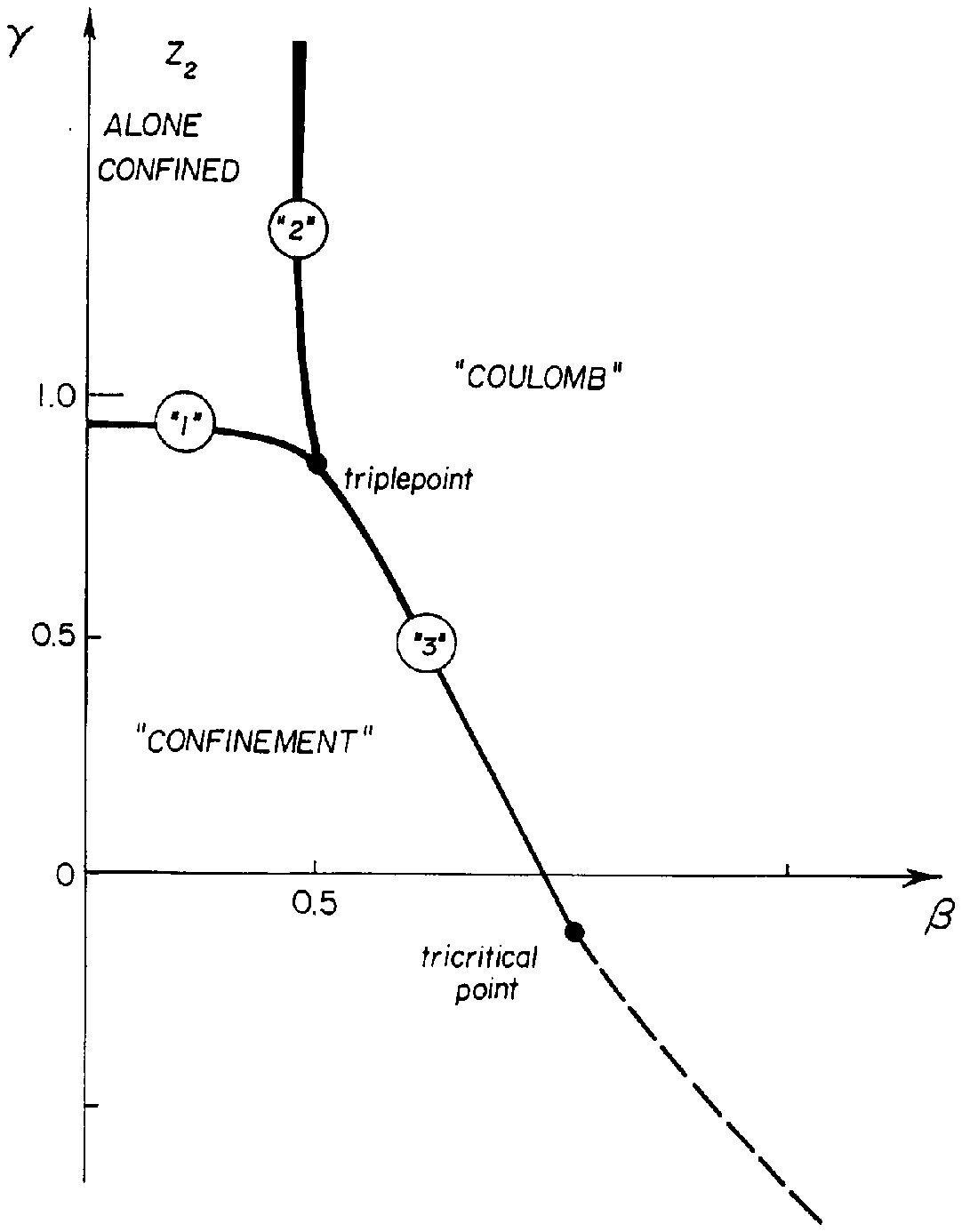}}
\caption{The phase diagram for $U(1)$ when the two-parameter
action is used. This type of action makes it possible to provoke the confinement
$Z_2$ (or $Z_3$) alone.}
\end{figure}
\newpage

\begin{figure}
\centerline{\epsfxsize=\textwidth \epsfbox{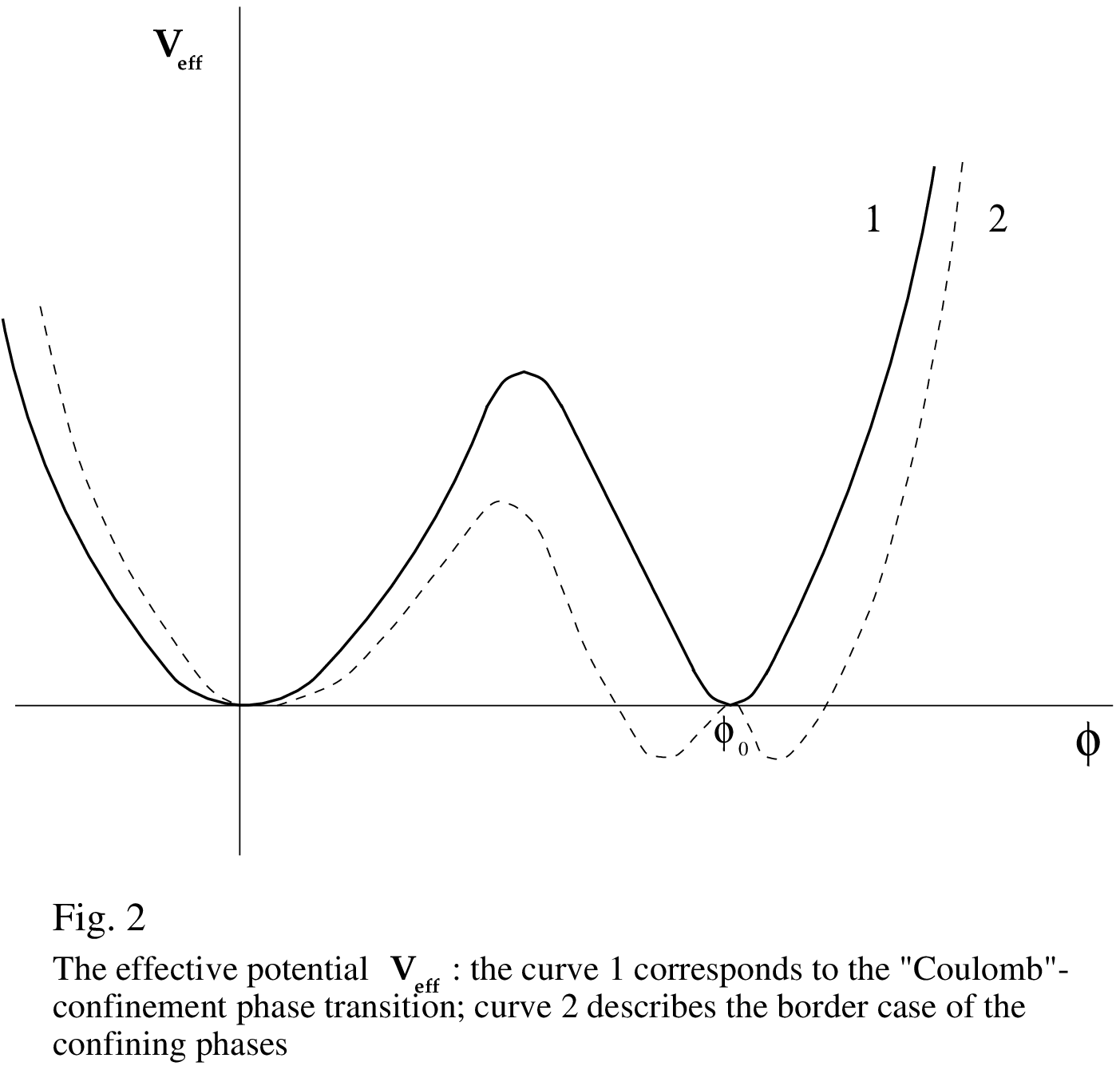}}
\end{figure}

\setcounter{figure}{2}

\newpage

\clearpage

\begin{figure}[t]
\bc
\centerline{\epsfxsize=\textwidth \epsfbox{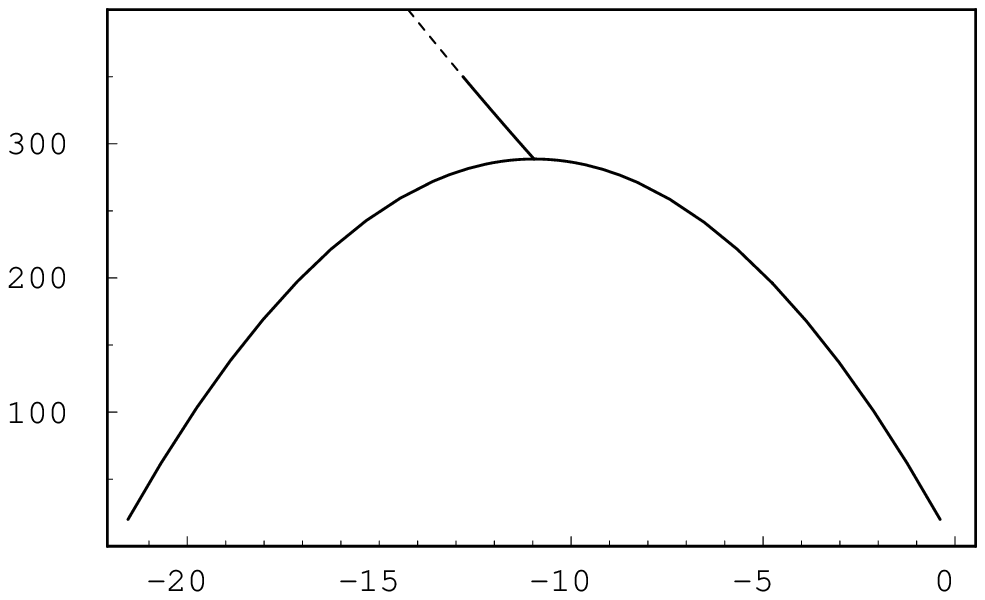}}

\bp(170,0)
%\put(12,148){\special{psfile=figpt3.eps hoffset=-90
%voffset=-400 hscale=170 vscale=220}}

\put(78,100){\bf Coulomb}
\put(88,90){\bf phase}
\put(-30,200){\Large\bf $V^{''}>0$}
\put(-30,180){\large\bf $\mbox{Confinement}_1$}
\put(-30,170){\large\bf phase}
\put(-72,207){\Large\bf $g^4$}

\put(235,0){\Large\bf $\lambda_{run}$}
\put(-72,30){\bf $0$}

\put(185,200){\Large\bf $V^{''}<0$}
\put(150,180){\large\bf $\mbox{Confinement}_2$}
\put(195,170){\large\bf phase}

\put(80,200){\Large\bf $1$}
\put(210,50){\Large\bf $2$}
\put(-20,50){\Large\bf $3$}
\put(100,170){\Large\bf $A$}

\put(100,32){\line(0,1){4}}
\put(100,41){\line(0,1){5}}
\put(100,51){\line(0,1){5}}
\put(100,61){\line(0,1){5}}
\put(100,71){\line(0,1){5}}
\put(100,81){\line(0,1){5}}
%\put(74,51){\line(0,1){5}}
%\put(74,61){\line(0,1){5}}
\put(100,111){\line(0,1){5}}
\put(100,121){\line(0,1){5}}
\put(100,131){\line(0,1){5}}
\put(100,141){\line(0,1){6}}
\put(100,151){\line(0,1){5}}
\put(100,161){\line(0,1){5}}
%\put(100,171){\line(0,1){5}}
%\put(100,181){\line(0,1){5}}
%\put(100,191){\line(0,1){5}}
%\put(74,21){\line(0,1){86}}
\put(97.5,40){\large\bf \llap{$\lambda_{run}^{(A)}$}}
\ep
\caption{Phase diagram ($\lambda_{run},g^4$) corresponding to the Higgsed
monopole model shows the existence of a triple point A
$\bigl(\lambda_{run}^{(A)}\approx -11,\;(g^{(A)})^2\approx
17\bigr)$}
\ec
\end{figure}


\newpage
\begin{thebibliography}{99}
\bibitem{1}
H.P.Nilles, Phys.Reports {\bf 110}, 1 (1984).
\bibitem{2}
P.Langacker, N.Polonsky, Phys.Rep. {\bf D47}, 4028 (1993).
\bibitem{3}
H.B.Nielsen, "Dual Strings. Fundamental of Quark Models", in:
{\it Proceedings of the XVII Scottish University Summer Scool in Physics},
St.Andrews, 1976, p.528.
\bibitem{4}
D.L.Bennett, H.B.Nielsen, I.Pi\^cek, Phys.Lett. {\bf B208}, 275 (1988).
\bibitem{5}
H.B.Nielsen, N.Brene, Phys.Lett. {\bf B233}, 399 (1989); Nucl.Phys.
{\bf B224}, 396 (1983).
\bibitem{6}
C.D.Froggatt, H.B.Nielsen, {\it Origin of Symmetries}, Singapore:
World Scientific, 1991.
\bibitem{7}
H.B.Nielsen and D.L.Bennett, "Fitting the Fine Structure Constants by
Critical Couplings and Integers",in: H.J.Kaiser, editor, {\it Proceedings
of the XXV International Symposium Ahrenshoop on the Theory of Elementary
Particles}, page 366. Institut fur Hochenergiephysik, Platanenallee 6,
D-O-1615 Zeuthen, Germany, Gosen, Sept.23-26 1991.
\bibitem{8}
D.L.Bennett and H.B.Nielsen, "Standard Model Couplings from Mean Field
Criticality at the Planck Scale and a Maximum Entropy Principle", in:
D.Axen, D.Bryman, and M.Comyn, editors,{\it Proceedings of the Vancouver
Meeting on Particles and Fields '91}, 18-22 August,
World Scientific Publishing Co., Singapore, 1992, page 857.
\bibitem{9}
D.L.Bennett and H.B.Nielsen, Int. J. Mod. Phys.{\bf A9}, 5155 (1994).
\bibitem{10}
L.V.Laperashvili, Phys.of Atom.Nucl.{\bf 57}, 471 (1994);ibid {\bf 59},
162 (1996).
\bibitem{11}
C.D.Froggatt, M.Gibson, H.B.Nielsen, D.J.Smith, "The Fermion Mass Problem
and the Anti-Grand Unification Model", in: {\it Proceedings of the
29th International Conference on High Energy Physics}, Vancouver,
Canada, 23--29 July, 1998; Int.J.Mod.Phys.{\bf A13}, 5037 (1998).
\bibitem{12}
L.V.Laperashvili, H.B.Nielsen, Mod.Phys.Lett.{\bf A12}, 73 (1997).
\bibitem{13}
C.D.Froggatt, L.V.Laperashvili, H.B.Nielsen, "SUSY or NOT SUSY: Anti-GUT's,
Critical Coupling Universality and Higgs--Top Masses", "SUSY98",
Oxford, 10-17 July 1998; hepwww.rl.ac.uk/susy98/.
\bibitem{14}
L.V.Laperashvili, H.B.Nielsen, "Multiple Point Principle
and Phase Transition in Gauge Theories", in:{\it Proceedings of the
International Workshop on "What Comes Beyond the Standard Model"},
Bled, Slovenia, 29 June - 9 July 1998; Ljubljana 1999, p.15.
\bibitem{15}
D.Zwanziger, Phys.Rev. {\bf D3}, 343 (1971).
\bibitem{16}
R.A.Brandt, F.Neri, D.Zwanziger, Phys.Rev.{\bf D19}, 1153 (1979).
\bibitem{17}
F.V.Gubarev, M.I.Polikarpov, V.I.Zakharov, Phys.Lett.{\bf B438}, 147 (1998).
\bibitem{18}
S.Coleman, E.Weinberg, Phys.Rev.{\bf D7}, 1888 (1973).
\bibitem{19}
M.Sher, Phys.Rept.{\bf 179}, 274 (1989).
\bibitem{20}
G.Bhanot, Nucl.Phys.{\bf B205}, 168 (1982); Phys.Rev.{\bf D24}, 461 (1981);\\
Nucl.Phys.{\bf B378} 633 (1992).
\bibitem{21}
J.Jersak, T.Neuhaus and P.M.Zerwas, Phys.Lett.{\bf B133} 103 (1983);\\
Nucl.Phys.{\bf B251}, 103 (1985).
\bibitem{22}
H.G.Everetz, T.Jersak, T.Neuhaus, P.M.Zerwas,
Nucl.Phys.{\bf B251}, 279 (1985).
\bibitem{23}
J.Jersak, T.Neuhaus, H.Pfeiffer, "Scaling Analysis of the Magnetic Monopole
Mass and Condensate in the Pure U(1) Lattice Gauge Theory",\\
hep-lat/9903034 v2, 7 April 1999.
\bibitem{24}
J.Jersak, C.B.Lang, T.Neuhaus, Phys.Rev.Lett. {\bf 77}, 1933 (1996);\\
Phys.Rev.{\bf D54}, 6909 (1996).
\bibitem{25}
J.Cox, W.Franzki, J.Jersak, C.B.Lang, T.Neuhaus, P.W.Stephenson, \\
Nucl.Phys.{\bf B499}, 371 (1997).
\bibitem{26}
J.Cox, W.Franzki, J.Jersak, C.B.Lang, T.Neuhaus, Nucl.Phys.{\bf B532},
315 (1998).
\bibitem{27}
J.Cox, J.Jersak, T.Neuhaus, P.W.Stephenson, A.Seyfried, H.Pfeiffer, \\
Nucl.Phys.{\bf B545}, 607 (1999).
\bibitem{28}
I.Campos, A.Cruz, A.Tarancon, Phys.Lett.{\bf B424}, 328 (1998);\\
Nucl.Phys.{\bf B528}, 325 (1998);
hep-lat/9711045, hep-lat/9803007, hep-lat/9808043.
\bibitem{29}
G.Damm, W.Kerler, Phys.Rev.{\bf D59}, 014510 (1999); hep-lat/9806036,\\
hep-lat/9808040.
\bibitem{30}
G.Arnold, T.Lippert, K.Shilling, Phys.Rev.{\bf D59}, 054509 (1999);
hep-lat/9809160.
\bibitem{31}
B.Klaus, C.Roiesnel, Phys.Rev.{\bf D58}, 114509 (1998); hep-lat/9801036.
\bibitem{32}
D.Horn, M.Karliner, E.Katznelson, S.Yankielowicz,\\
Phys.Lett.{\bf B113}, 258 (1982).
\bibitem{33}
D.Horn, E.Katznelson, Phys.Lett.{\bf B121}, 349 (1983).
\bibitem{34}
J.L.Cardy, Nucl.Phys.{\bf B170}, 369 (1980).
\bibitem{35}
N.N.Bogoljubov, D.V.Shirkov, Doklady AN SSSR (Reports of AS USSR),\\
\un{103}(1955)203; ibid {\bf 103}, 391 (1955); JETP,{\bf 30}, 77 (1956).
\bibitem{36}
L.D.Landau, A.A.Abrikosov, I.M.Khalatnikov, Doklady AN SSSR \\
(Reports of AS USSR),{\bf 95}, 773 (1954); ibid {\bf 95}, 1177 (1954).
\bibitem{37}
L.V.Laperashvili, H.B.Nielsen, "Dirac Relation and Renormalization
Group Equations for Electric and Magnetic Fine Structure Constants",
to be published.
\bibitem{38}
P.A.M.Dirac, Proc.Roy.Soc.{\bf A33}, 60 (1931).
\bibitem{39}
P.Di Vecchia, "Duality in Supersymmetric Gauge Theories", Surveys in
High Energy Physics, Vol.{\bf 10}, 119 (1997); hep-th/9608090;\\
"Duality in N=2,4 Supersymmetric Gauge Theories", preprint Nordita
98/11-HE; hep-th/9803026 v2.
\bibitem{40}
J.Schwinger, Phys.Rev.{\bf 144}, 1087 (1966);ibid {\bf 151}, 1048,1055 (1966);
ibid {\bf 173}, 1536 (1968); \\
Science {\bf 165}, 757 (1969); ibid {\bf 166}, 690 (1969).

\end{thebibliography}
\end{document}